\documentclass[twocolumn]{aastex62}
\usepackage{graphicx}
\usepackage[normalem]{ulem}
\newcommand\redout{\bgroup\markoverwith
{\textcolor{red}{\rule[.5ex]{2pt}{0.4pt}}}\ULon}
\usepackage{xcolor}
\usepackage{threeparttable}
\usepackage{amsmath}

\newcommand{\msun}{$M_{\odot}$}
\newcommand{\msuneq}{M_{\odot}}
\newcommand{\hone}{\textsc{H\,i}}
\newcommand{\lya}{Ly$\alpha$}

\newcommand{\rhorhom}{$\rho_m /\langle \rho_m \rangle$}
\newcommand{\rhom}{$\rho_m$}
\newcommand{\rhophys}{$\rho_{\rm Phys}$}
\newcommand{\ngal}{37,662}
\newcommand{\kms}{km~s$^{-1}$}

\graphicspath{{./}{figures/}}

\accepted{to ApJL 01/23/2020}

\shorttitle{Revealing the Dark Threads of the Cosmic Web}
\shortauthors{Burchett et al.}

\begin{document}

\title{Revealing the Dark Threads of the Cosmic Web}

\correspondingauthor{Joseph N. Burchett}
\email{burchett@ucolick.org}

\author{Joseph N. Burchett}
\affil{Department of Astronomy \& Astrophysics, University of California, 1156 High Street, Santa Cruz, CA 95064, USA}
\author{Oskar Elek} 
\affiliation{Department of Computational Media, University of California, 1156 High Street, Santa Cruz, CA 95064, USA}
\author{Nicolas Tejos}
\affiliation{Pontificia Universidad Cat\'olica de Valpara\'iso}
\author[0000-0002-7738-6875]{J. Xavier Prochaska}
\affil{Department of Astronomy \& Astrophysics, University of California, 1156 High Street, Santa Cruz, CA 95064, USA}
\affiliation{Kavli Institute for the Physics and Mathematics of the Universe (Kavli IPMU),
5-1-5 Kashiwanoha, Kashiwa, 277-8583, Japan}
\author{Todd M. Tripp}
\affiliation{University of Massachusetts -- Amherst}
\author{Rongmon Bordoloi}
\affiliation{North Carolina State University}
\author{Angus G. Forbes}
\affiliation{Department of Computational Media, University of California, 1156 High Street, Santa Cruz, CA 95064, USA}

\begin{abstract}
Modern cosmology predicts that matter in our Universe has assembled
today into a vast network of filamentary structures colloquially
termed the Cosmic Web.
Because this matter is either electromagnetically invisible (i.e., dark)
or too diffuse to image in emission, tests of this cosmic
web paradigm are limited.  Wide-field surveys do reveal
web-like structures in the galaxy distribution,  but these luminous galaxies represent less
than 10\% of baryonic matter.  Statistics of absorption by the
intergalactic medium (IGM)
via spectroscopy of distant quasars support the model yet
have not conclusively tied the diffuse IGM to the web.
Here, we report on a new method inspired by the \textit{Physarum polycephalum} slime mold that is able to infer the density field of the Cosmic Web from galaxy surveys.
Applying our technique
to galaxy and absorption-line surveys of the local Universe, we demonstrate that the bulk of the IGM indeed resides in the Cosmic Web.  
 From the outskirts of Cosmic Web filaments, at approximately the cosmic mean matter density (\rhom) and $\sim 5$ virial radii from nearby galaxies, we detect an increasing \hone\ absorption signature towards higher densities and the circumgalactic medium, to $\sim 200$\rhom.  However, the absorption is suppressed within the densest environments, 
suggesting shock-heating and ionization deep within filaments and/or feedback processes within galaxies.
\end{abstract}


\section{Introduction}

After decades of study, the $\Lambda$ Cold Dark Matter ($\Lambda$CDM) cosmological framework is now significantly refined, and many basic parameters of this theory are constrained by observations.  At this stage, investigations of many key cosmology tests and galaxy evolution questions require computationally intensive large-scale simulations that attempt to incorporate a wide variety of physical processes with adequate spatial and mass resolution to capture essential behaviors of galaxies and the universe.  These cosmological simulations are highly complex, generally relying on varying assumptions or `prescriptions' for dealing with these unresolved physical processes hydrodynamically \citep[e.g,][]{Oppenheimer:2006uq,Schaye:2015yg}.  In principle, the intergalactic medium (IGM), which should primarily involve hydrogen ionization and recombination, provides one of the most straightforward tests of the simulations \citep{Dave:1999_lyaForest}.  However, the IGM has extremely low densities and is therefore challenging to detect.  Moreover, galaxies are surrounded by a gaseous circumgalactic medium \citep[CGM; e.g.,][for a review]{Bergeron:1991qy,Lanzetta:1995rt,Keeney:2013nx,Tumlinson:2017aa} that is affected by a more complicated mixture of galactic inflows, outflows, and energy `feedback', and it is not immediately clear how to distinguish hydrogen in the CGM from the intergalactic hydrogen so that the simulations can be tested in the relatively simple regime of the IGM.

\begin{figure*}[t!]
\includegraphics[width=\textwidth]{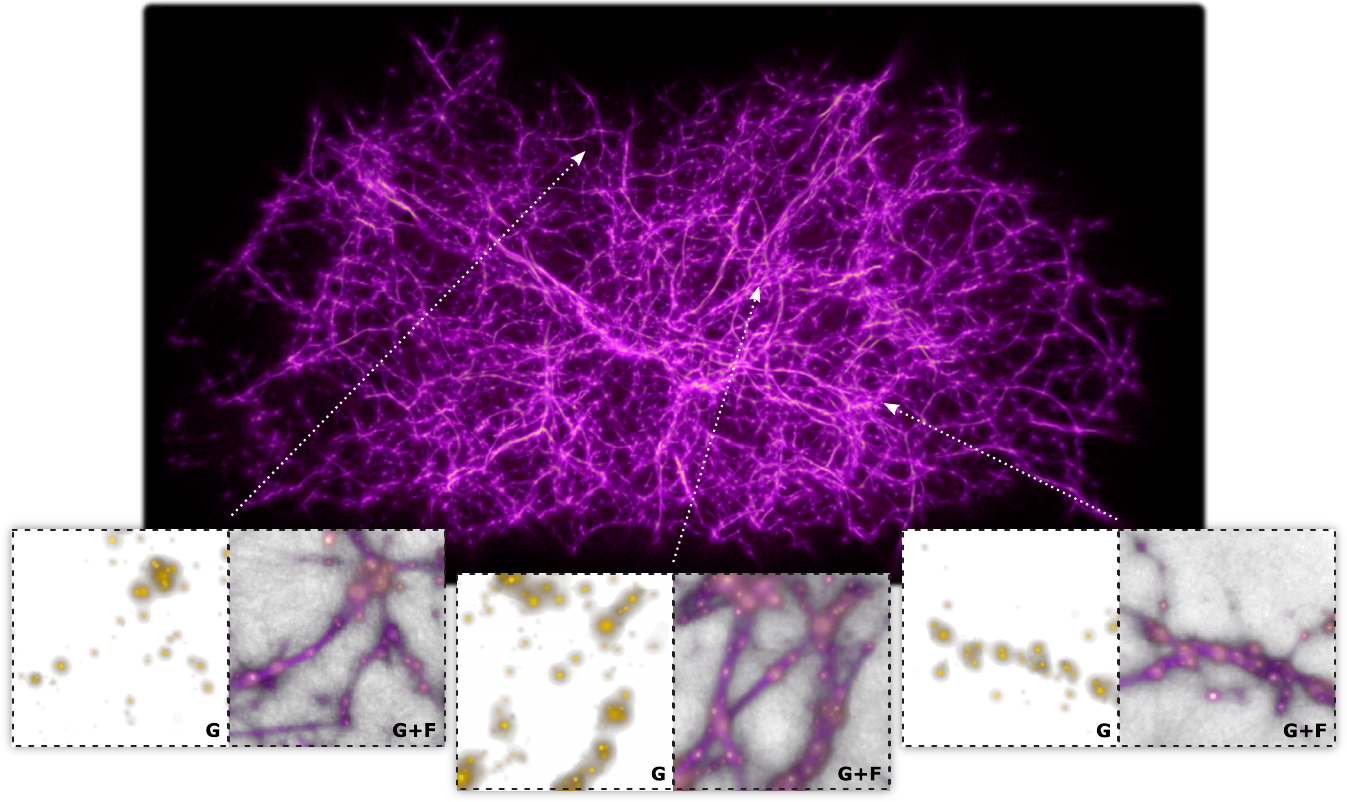}
\caption{\footnotesize
\textbf{MCPM Cosmic Web reconstruction using \ngal\ galaxies from SDSS.}
Top: Large-scale visualization of the emergent structure identified by our Monte Carlo Physarum Machine (MCPM) algorithm.
This intricate filamentary network is reconstructed given only the SDSS galaxy coordinates, redshifts, and masses.
Bottom: Three individual regions showing the underlying SDSS galaxies (G) and the superimposed MCPM filament density field (G+F).
}
\label{fig:teaser_sdss}
\end{figure*}

As the most sensitive method for detecting the highly diffuse gas in the Universe, absorption line spectroscopy has been the primary technique for probing the CGM and IGM.  Generally, the technique involves using a bright background source such as a quasi-stellar object (QSO) or a galaxy to probe the halo of a foreground object projected near the line of sight. On scales beyond the CGM, the emergent Cosmic Web \citep{Bond:1996aa} -- generically predicted in cosmological models of a CDM Universe -- should be permeated by low-density IGM gas \citep{Cen:1994aa}. Efforts to correlate absorption signatures with galaxies on IGM scales have shown that gas and galaxies are indeed correlated over Megaparsec-scale distances \citep{Morris:1993kq,Chen:2009vn,Tejos:2012lr,Tejos:2014fu,Prochaska:2019aa}, and \hone\ absorption tomography using multiple sightlines probing a single filament suggests a stronger absorption signal near the center of that filament \citep{Wakker:2015rf}.  Under the assumption that cosmic filaments connect massive galaxy clusters, it has also been shown that warm-hot (10$^{5-6}$\,K) gas is enhanced within filaments \citep{Tejos:2016qv,Pessa:2018_filaments}, indicating that the IGM is heated in these over-dense environments.
However, it remains to be demonstrated over a cosmologically representative volume that the IGM absorbers routinely detected in QSO spectroscopy indeed trace the {\it same} large-scale Cosmic Web structure evident in galaxy redshift surveys \citep{deLapparent:1986_sliceUniverse}.   

The principal challenge is identifying and quantifying the large-scale structure of the Cosmic Web, particularly in regions where the most diffuse gas is purported to exist.  
Several groups have developed automatic methods to identify large scale structures and/or derive the underlying density fields \citep[e.g.,][]{Aragon-Calvo:2007qy,Cautun:2013aa_nexus,Sousbie:2011ab_disperse2,Tempel:2014aa,Chen:2016aa_filamentCatalogue,Libeskind:2018aa}.   The difficulty of the problem for all of these methods is compounded by the lack of a `ground truth' solution for large-scale structures in the observed Universe.  These methods have been fruitfully applied to galaxy survey data to show that morphology, star formation rate, and cold gas content depend not only on their halo masses but also on a galaxy's location within the large scale structure \citep[e.g.,][]{Stark:2016lr,Chen:2017_filaments,Kuutma:2017aa}.  However, associating the IGM absorption with this underlying structure requires key features that are often missing from such methods.  First, the cosmic web density values must be defined at points in the survey volume on scales of several Megaparsecs away from galaxies, where the QSO sightlines probe low density regions of the IGM.  Second, the environment must be quantified across a wide dynamic range of scale, from small scales of galaxy groups to the larger scales of cosmic web filaments.  The former  generally excludes methods such as nearest-neighbor distances, and the latter generally excludes those based on purely geometrical filament identification.  

Here, we present and apply an alternative novel approach inspired by the \textit{Physarum polycephalum} slime mold, which attempts to satisfy both requirements simultaneously, as shown in Figure \ref{fig:teaser_sdss}.  Our model implicitly traces the Cosmic Web structure by finding efficient continuous pathways between galaxies, which are known to be preferentially distributed along filaments.

%%%%%%%%%%%%%%%%%%%%%%%%%%%%%%%%%%%%%%%%%%%%%%%%%%%%%%%%%%%%%%%%%%%%%%%%%%%%%%%%%%%%%%
\section{Methods}

\subsection{Reconstructing the Cosmic Web}

A key difficulty in identifying Cosmic Web structures such as filaments is the heavily under-constrained nature of the problem: how does one reconstruct a physically plausible 3D structure out of sparse and often heterogeneous observations?
We approach this challenge from the perspective of finding a meaningful interpolator for such sparse data---put in other words, finding a generative model conditioned by the data and informed by an appropriate statistical prior.

Our generative model is heavily inspired by the work of \citet{Jones:2010aa}, who proposed an agent-based algorithm to mimic the growth and development of the \textit{Physarum polycephalum} `slime mold'.
This model, a kind of virtual Physarum machine, is a computational counterpart to analog Physarum machines.
The latter have been applied over the last 20 years to solve mazes \citep{Nakagaki:2000}, shortest path finding \citep{Nakagaki:2001}, transportation network design \citep{Tero:2010}, and many other optimal transport and spatial organization problems \citep{Adamatzky:2010}, some of which are NP-hard (i.e., among the most computationally difficult).

Jones's model \citep{Jones:2010aa} finds an optimal transport network via a feedback loop between a swarm of discrete agents (analogous to slime mold cells) and a continuous lattice of `chemo-attractant' deposited by the agents and the input data.  Within the model domain, the agents move along paths determined by the presence of the attractant and emit their own deposit at each time step.  We have adapted this model to three dimensions and performed a critical modification: a \textit{probabilistic sampling and selection} of the agents' spatial trajectories.  In its original form, Jones's algorithm imposes that agents' trajectories maximize the deposit values encountered; our version samples possible paths probabilistically, such that paths toward smaller deposits may still be traversed, although less frequently. This stochastic process formulation enables us to frame the extracted 3D network as a probability density of the presence of filamentary structures, rather than an absolute maximum likelihood solution.
We call the resulting framework the \textit{Monte-Carlo Physarum Machine} (MCPM, see Appendix).

\begin{figure}[t!]
    \includegraphics[width=0.49\textwidth]{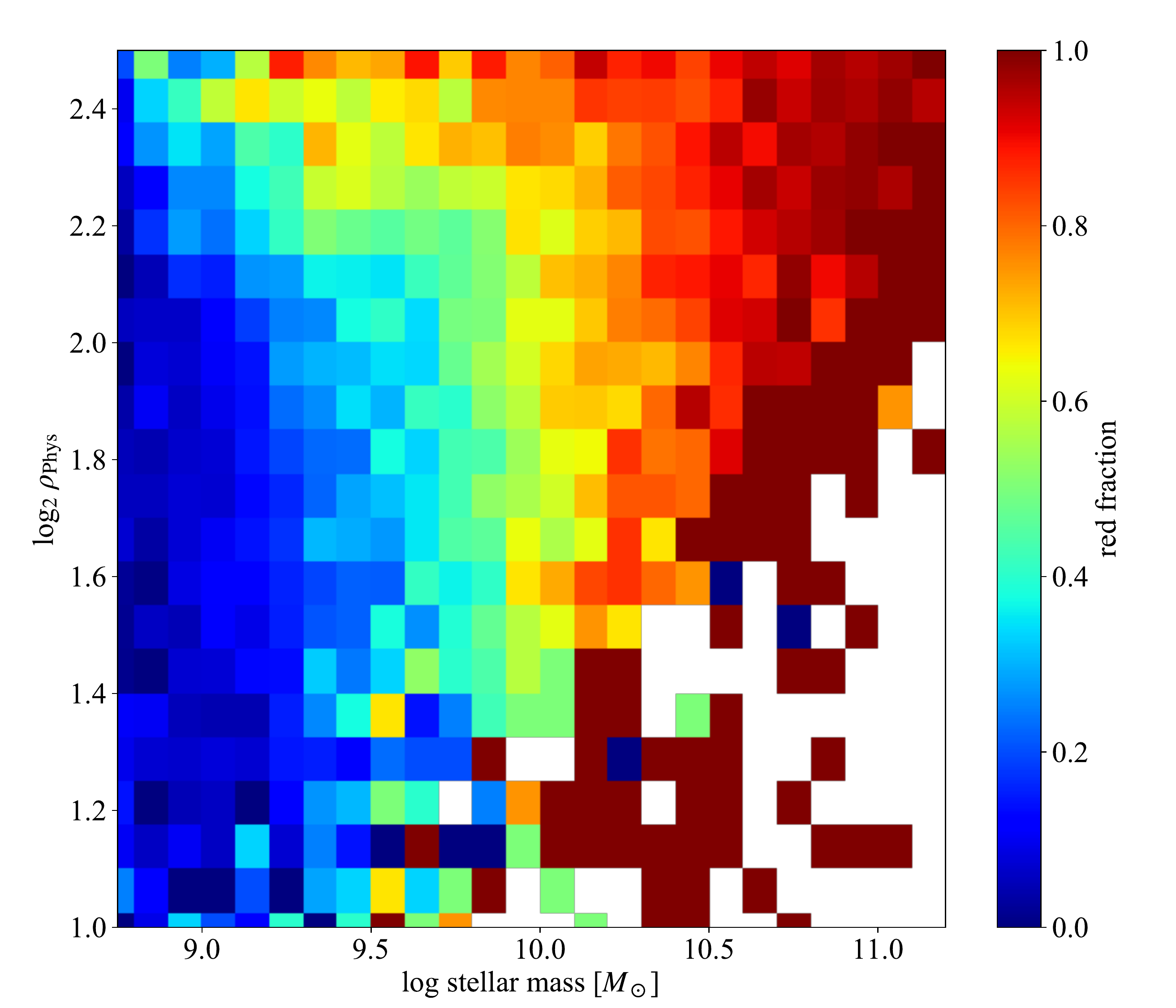}
      \caption{\footnotesize \textbf{The dependence of star formation activity on galaxy environment and stellar mass for the galaxies within our SDSS volume.}  The color coding denotes the fraction of `red' galaxies in the population within each mass/environment bin, where the environmental density (\rhophys) is determined from our MCPM Cosmic Web reconstruction algorithm.  A comparison with Figure 6 of \citet{Peng:2010tg} shows a similarly increasing red fraction as functions of both mass at fixed density and density at fixed mass.}
    \label{fig:massEnviroQuenching}
\end{figure}

\begin{figure*}[t!]
     \centering
    \includegraphics[width=0.7\linewidth]{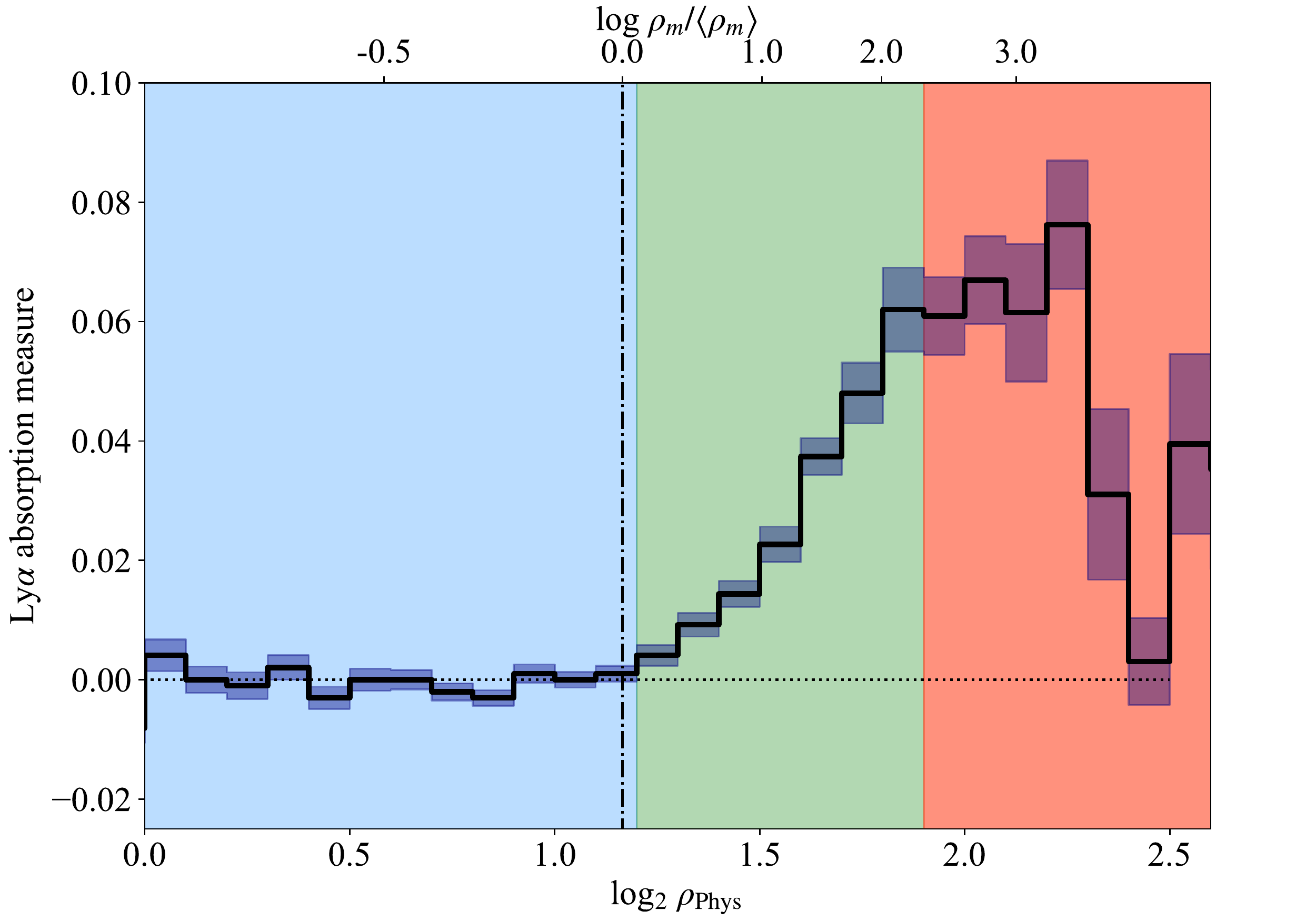}
    \vspace{-2mm}
    \begin{flushleft}
    \caption{\footnotesize \textbf{Median H I absorption as a function of Cosmic Web density.}  The lower horizontal axis represents density (trace) values from our MCPM model while the upper axis shows the corresponding density in terms of the $z=0$ cosmic mean matter density (see Appendix).  Bootstrap errors in the median are shaded about the black median absorption measure curve. Three density regimes are highlighted according to the absorption measure: low density (blue), where the signal is essentially flat with the horizontal dotted line representing a linear fit to those values; intermediate density (green), where the absorption signal monotonically increases with increasing density; and high density (red), where the absorption trend abruptly halts and even reverses. \label{fig:absorptionSignal}}
    \end{flushleft}
    \vspace{-4mm}
\end{figure*}

We have applied the MCPM method to a sample of \ngal\ spectroscopically observed galaxies from the Sloan Digital Sky Survey (SDSS) at redshifts $z=0.0138-0.0318$.  This sample exhibits a number of Cosmic Web filaments as well as massive galaxy clusters (including Coma) and voids, clearly evident from our 3D visualization tool \citep{Burchett:2019bb}; the volume is also pierced by $>500$ QSO sightlines observed with the Cosmic Origins Spectrograph (COS) aboard the Hubble Space Telescope (HST) and that cover the \hone\ \lya\ transition over these redshifts.  Figure~\ref{fig:teaser_sdss} shows our resulting Cosmic Web reconstruction, where the galaxies effectively serve as `food' sources for a swarm of virtual `slime mold' agents released into a 3D space defined by the celestial coordinates of each galaxy and the luminosity distance implied by each galaxy's redshift as the radial coordinate.  The agents continually move through space and eventually reach an equilibrium state, tracing an approximate optimal transport network from galaxy to galaxy given a small number of model parameters (see Appendix).

Essential to the equilibration of the model is a \textit{deposit} emitted by both the galaxies and the agents at each time step; this deposit is sensed by nearby agents, enabling the creation and reinforcement of tentative pathways.  Visualized in the top panel of Figure~\ref{fig:teaser_sdss}, for example, is the \textit{trace}, which records the aggregate agents' trajectories, and we in turn leverage this trace as a proxy of density values at each point in the 3D space. While the galaxies are directly observed, the purpose of the âagentsâ is to model the distribution of the (unobserved) dark matter that molds the Cosmic Web; the âtraceâ is the map of web filaments and structures found in the converged agent and galaxy distribution.

\subsection{Validating the Model with Galaxy Populations}
The well-established correlations between galactic star formation and local environment \citep{Balogh:1999xy,Kauffmann:2004ai,Peng:2010tg,Rasmussen:2012aa,Treyer:2018aa} and stellar mass \citep{Baldry:2004qy} provide a natural test for our density reconstruction (see the Appendix for additional model tests).  Crossmatching the input galaxy catalog with our MCPM density field, we present the two-dimensional locus of the `red' galaxy fraction (SDSS color $g-i>0.8$) as a function of local environmental density and stellar mass in Figure~\ref{fig:massEnviroQuenching}.  At any fixed stellar mass, the red fraction increases with increasing environmental density; at fixed density, the red fraction increases with stellar mass.  A comparison with Figure 6 of \citet{Peng:2010tg} shows qualitatively similar dependences of galaxy quenching on both mass and environment.  Our results are also in agreement with more recent studies \citep{Kuutma:2017aa} that report decreasing star formation rates for galaxies residing in the denser regions of filaments.

%%%%%%%%%%%%%%%%%%%%%%%%%%%%%%%%%%%%%%%%%%%%%%%%%%%%%%%%%%%%%%%%%%%%%%%%%
\subsection{H I in the Cosmic Web}
 A key strength of the Physarum model is that it provides a density value (\rhophys) for every point in the three-dimensional space (i.e., not only at the locations of galaxies), and each pixel of the HST/COS spectra skewering the volume can be assigned a corresponding density value. We then measure the absorption signal from a given spectral transition as a function of the local density of the Cosmic Web.  Two of the three coordinates needed to localize each voxel within the Physarum data cube are provided by the celestial coordinates of each sightline, and the third is the luminosity distance of each would-be redshift assuming that an \hone\ \lya\ line falls at the wavelength of that pixel. Figure~\ref{fig:absorptionSignal} shows the \hone\ absorption signal as a function of the local density, where we have binned the HST/COS spectral pixels by the Physarum trace values probed by these spectra and calculated the median normalized flux for pixels in each bin.  The absorption signal is expressed as a quantity similar to the optical depth: 
\begin{equation}
\tau = {\rm ln} \left(\frac{f_b}{f_i} \right)
\end{equation}
where $\tau$ is the absorption measure, $f_b$ is the normalized baseline flux obtained from a linear fit to the absorption measure values at log$_2$ \rhophys $< 0.8$ (see Appendix), and $f_i$ is the normalized flux in a given pixel. The blue shading along the curve in Figure \ref{fig:absorptionSignal} indicates 1$\sigma$ uncertainties on the median from bootstrap resampling. To calibrate the Physarum trace values \rhophys\ to cosmologically meaningful quantities, we employed the Bolshoi-Planck cosmological simulation \citep{Klypin:2016aa}.  We fitted the MCPM to a halo catalog \citep{Behroozi:2013aa} from the simulation and mapped the \rhophys\ values from this fit to the dark matter density field (in units of the cosmic mean matter density \rhorhom) from the particle data; the corresponding \rhorhom\ for the SDSS \rhophys\ values are labelled on the upper axis of Figure~\ref{fig:absorptionSignal}.

\section{Results}
\begin{figure*}[t!]
\centering 
\includegraphics[width=0.8\textwidth]{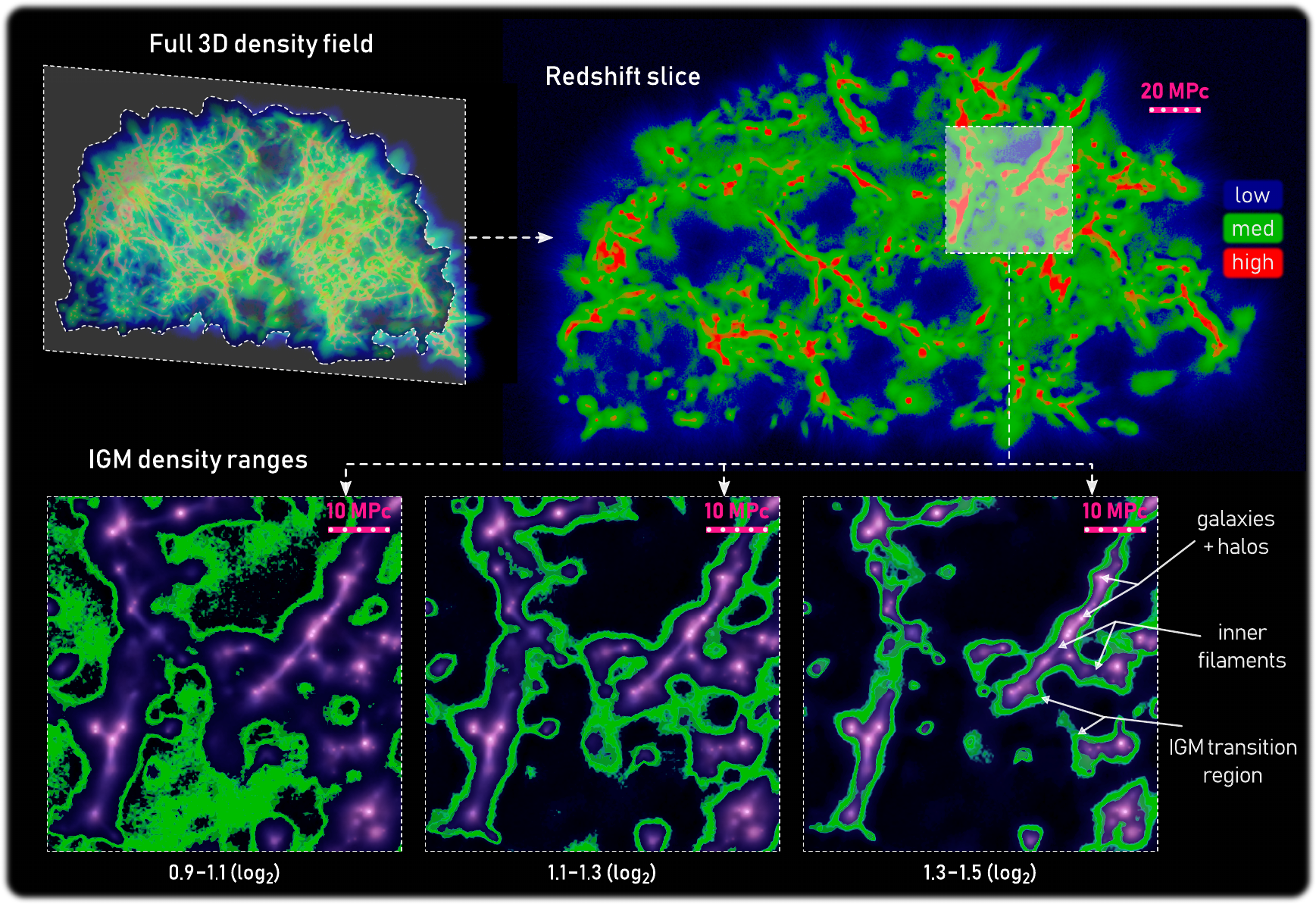}
\caption{\footnotesize
\textbf{A detailed inspection of the IGM transition region.} (Top-left) A heatmap visualization of the full density field in a 3\,Mpc thick slice from our MCPM fit to the SDSS galaxy sample.  (Top-right) The same 3\,Mpc slice segmented and color-coded by the three density regimes: low-, moderate-, and high-densities are colored blue, green, and red, respectively, as in Figure \ref{fig:absorptionSignal}.  (Bottom) Highlighting three bins in density bracketing the low-moderate density transition shown in Figure \ref{fig:absorptionSignal} within a single spatial region. The deposits of the galaxies input to the MCPM algorithm are shown in magenta, and the MCPM trace is colored green in each panel with corresponding log$_2$ \rhophys\ values indicated below each panel.  The left panel shows that the density regime just before the onset of the \hone\ absorption signal includes regions within putative voids, whereas the center and right panels demonstrate that the onset of significant \hone\ absorption occurs in the outskirts of filaments.
}
\label{fig:bracketing}
\end{figure*}

Three key features emerge from Figure~\ref{fig:absorptionSignal}: 1) the `low-density' regime at Physarum model densities log$_2$~$\rho_{\rm Phys} \lesssim 1.2$ (shaded in blue), where we show few statistically significant deviations from the baseline flux value; 2) the `moderate-density' regime at $1.2 < {\rm log_2}~\rho_{\rm Phys} < 1.9$ (shaded in green), where an absorption signal steadily increases; and 3) the `high-density' regime at log$_2$~$\rho_{\rm Phys} > 1.9$ (shaded in red), where the absorption signal no longer increases and shows an \textsl{abrupt attenuation} at log$_2$~$\rho_{\rm Phys} \sim 2.3$.  The moderate density regime sets in at approximately the mean matter density of the Universe.  Figure~\ref{fig:bracketing} highlights the low-intermediate density transition, demonstrating that our absorption signal is indeed arising from the far outskirts of filaments and revealing the dark and diffuse fringes of the Cosmic Web.  

Previous studies \citep[e.g.,][]{Morris:1993kq,Tripp:1998kq,Tejos:2012lr,Penton:2002aa,Stocke:2007kq,Wakker:2009fr,Prochaska:2011aa} have suggested that weak \hone\ absorbers (column densities of $N$(\hone) $< 10^{13.5}$\,cm$^{-2}$) may indeed arise from highly underdense regions of the Cosmic Web (voids). However, the definitions of voids vary widely and are typically expressed in terms of the distance to nearby galaxies, e.g., $>$1.4\,Mpc to the nearest L* galaxy \citep{Stocke:2007kq}. Our method attempts to define structures in terms of the underlying density field. However, for comparison, we crossmatched our sightlines with potentially associated galaxies. In the density bin where our detected signal sets in (near $\langle \rho_m \rangle$), we find that the \lya\ absorbing spectral regions are generally $>2.5$\,Mpc from the nearest L* galaxy within $\pm 500$\,\kms.  We posit that some fraction of the few void absorbers previously reported may indeed be outliers in the distribution of \hone\ in underdense regions and represent rare large density perturbations uncorrelated with Cosmic Web filaments traced by galaxies \citep{Tejos:2012lr,Tejos:2014fu}. However, the monotonic rise of the absorption signal in Figure~\ref{fig:absorptionSignal} through the intermediate density regime suggests that weak \lya\ forest absorbers trace the outskirts of filaments, as also shown in Figure~\ref{fig:bracketing}, smoothly increasing in strength with the ambient density field as would be expected from results of hydrodynamical simulations \citep[e.g.,][]{Dave:2010qy}. 
    
In Figure~\ref{fig:absorptionSignal}, the absorption measure in the moderate-density regime increases with local density until it abruptly halts near log$_2$ \rhophys\ $= 1.9$, corresponding to \rhorhom\ $\sim 200$.  In the regime where the absorption measure is steadily increasing (green shaded region in Figure \ref{fig:absorptionSignal}), the hydrogen absorption segues from arising in truly intergalactic material that has no physical connection with any nearby galaxies to circumgalactic material that plausibly interacts with the closest galaxies. Crossmatching sightlines probing this regime with nearby galaxies suggests that the high-density region sets in at 
$\approx 2$~virial radii from nearby galaxies.  Here, galaxy feedback and other hydrodynamical effects are likely modulating gas physical conditions within the circumgalactic medium (CGM), heralding a physically distinct regime from the surrounding IGM.  The CGM as traced by \hone\ and heavy element ions is heavily dependent on host galaxy properties \citep{Tumlinson:2011kx,Bordoloi:2011uq,Tumlinson:2013cr,Lan:2014aa,Johnson:2015qv,Burchett:2016aa,Nielsen:2013ab}, such as mass, star formation rate, and environment.  The same processes reflected in these CGM tracers plausibly impart the sudden deviation from the intermediate density absorption trend.  

Within the high-density regime (shaded red in Figure~\ref{fig:bracketing}), the \hone\ absorption remarkably decreases at log$_2$ \rhophys\ $\sim 2.3$.  This density regime includes galaxy clusters and the inner regions (or `spines') deep within filaments.  In addition to the circumgalactic effects mentioned above, additional environmental factors are likely at play here.  First, within galaxy clusters, \hone\ absorption is highly suppressed, even in sightlines probing near galaxies \citep{Yoon:2013kq,Yoon:2017aa,Burchett:2018aa,Butsky:2019aa}.  The intracluster medium (ICM) is mostly composed of hot $>10^6$\,K gas, for which the neutral hydrogen fraction is extremely low; furthermore, the cool gas abundant in the CGM of galaxies outside clusters can be readily stripped upon falling into the ICM \citep{Bahe:2013yu,Tonnesen:2007yq,Zinger:2018aa}. Second, the filamentary environment itself is predicted to heat the intergalactic gas, which in turn will further ionize the gas and/or broaden the \hone\ absorption profiles, effectively reducing the absorption strength per pixel \citep{Richter:2006aa,Richter:2008aa,Tepper-Garcia:2012aa,Tonnesen:2017aa,Tejos:2016qv,Wakker:2015rf}.  An additional effect may occur in the moderate- and high-density regimes arising from high-peculiar velocity flows relative to  the galaxies we are using to trace the large-scale structure. This may redistribute flux to pixels attributed to higher or lower density regimes than that of its physical origin.

By employing the MCPM Cosmic Web reconstruction technique, we have characterized the neutral hydrogen content in the low-redshift Universe over four orders of magnitude in matter density. We have shown that \hone\ absorption from the IGM tracks the large scale structure of the Universe and smoothly increases from the outskirts of cosmic web filaments toward higher densities until processes in the CGM, ICM, and dense cores of filaments suppress the signal.  Regarding the application of MCPM (rooted in biology) within an astrophysical context, we emphasize that the Physarum behaviors modeled by MCPM do not attempt to model the physical origins of structure formation, i.e., gravity and dark energy.  Rather, the outcomes of these disparate modeled processes are related by their formation of optimized network structures \citep{Adamatzky:2010,Hong:2016_networks}.  Furthermore, the agents in our model follow trajectories throughout the structure, although these trajectories will differ from those traveled by galaxies in Cosmic Web filaments and galaxy clusters. However, we are analyzing these agent trajectories to glean further insight about the cosmic web structure (e.g., filament identification) for a subsequent publication. Future applications of MCPM will more deeply examine the intimate connections galaxies have with their large-scale environments and the resulting symbiosis with their gaseous surroundings.

\section*{Acknowledgements} We are grateful to Jan Ivaneck\'{y}, who authored an early prototype of the simulation code and remained a source of valuable technical advice, and Daisuke Nagai, who provided key helpful suggestions.
The simulation art of Sage Jenson inspired the idea of possible connection between the morphology of Physarum Polycephalum and the intergalactic medium structure.
Bolshoi-Planck simulation data were kindly provided by Joel Primack and Doug Hellinger.
Support for this work was provided by NASA through grant number HST-AR 15009 from the Space Telescope Science Institute, which is operated by AURA, Inc., under NASA contract NAS 5-26555.

%%%%%%%%%%%%%%%%%%%%%%%%%%%%%%%%%%%%%%%%%%%%%%%%%%%%%%%%%%%%%%%%%%%%
\section*{Appendix}

\subsection*{Observations and simulation data}
Our analysis employs two key observational datasets: 1) the NASA/Sloan Atlas (NSA), a value-added catalog of SDSS photometry and spectroscopy and 2) the Hubble Spectroscopic Legacy Archive \citep[HSLA;][]{Peeples:2017aa}, a database of reduced and coadded spectra from the Cosmic Origins Spectrograph (COS) aboard HST. We estimated star formation rates, halo masses, and virial radii for all galaxies within the NSA to accompany the stellar masses provided in the original catalog \citep{Burchett:2016aa}. We focus our analysis on a volume including a redshift range of $z=0.0138-0.0318$ over the spring SDSS footprint.  This sample was chosen to feature several attributes: a) a rich diversity in large scale structures including galaxy clusters, filaments, and voids; b) a high spectroscopic completeness to faint dwarf galaxies; and c) a large number of sightlines to background QSOs with HST archival spectra.  Indeed, the volume we analyze includes five Abell clusters \citep{Abell:1965fk}, including Coma, and several galaxy filaments apparent from visual inspection within SDSS (Figure \ref{fig:teaser_sdss}).  Our redshift slice, centered on Coma's brightest cluster galaxies, includes approximately $\pm3$ times the velocity dispersion of Coma \citep{Fuller:2016fj} and is spectroscopically complete down to 0.05\,$L*$ \citep[assuming the nominal magnitude limit of SDSS galaxy spectra;][]{Strauss:2002fk}.  Because the HSLA does not include QSO redshifts within the main database, we crossmatched all HSLA QSOs with the \textit{Million Quasars Catalog} \citep{Flesch:2015jk} v5.2 to obtain their redshifts. We then selected those QSOs with sightlines within the survey volume and with $z > 0.0318 + \Delta v / c$, for $\Delta v = 5000$\,\kms\ and where $c$ is the speed of light, to select sightlines probing the full volume and mitigate confusing foreground absorption with outflow signatures of distant QSOs \citep{Tripp:2008lr}. For the 346 QSO spectra meeting the criteria above and with S/N $>3$, we fit continua over the wavelength range covering \lya\ 1215.67\,\AA\ at $z=0.0138-0.0318$ using an automated procedure, which iteratively fits series of Legendre polynomials to the data while sigma clipping to reject pixels likely due to physical absorption features from the fit. The order of the polynomial series fitted is determined by an F-test.

We ultimately identify large scale structure and quantify the local density using our Monte Carlo Physarum Machine (MCPM) slime mold model.  To calibrate the density values extracted from this Physarum model with cosmologically meaningful values, we employ the dark matter only Bolshoi-Planck (BP) cosmological simulation \citep{Klypin:2016aa,Roca-Fabrega:2019_cgm}. The BP simulation volume comprises (250\,$h^{-1}$ Mpc)${^3}$ and contains $2048^3$ particles; we utilize the $z=0$ snapshot. Conveniently, the physical dimensions of the simulation exceed the luminosity distances of the most distant galaxies in our sample.  Therefore, we truncated the simulation data to a width of approximately 70\,Mpc in one direction centered on the median luminosity distance of our galaxy sample, which would ease the computational load when fitting the BP data with the MCPM algorithm.  Specifically, we used two BP data products in our analysis: a halo catalog produced using the Rockstar algorithm \citep{Behroozi:2013aa} and a datacube containing the dark matter density field smoothed over 0.25\,Mpc scales \citep{Goh:2019_haloCosmicWeb,Lee:2017_halosDensity}.

%%%%%%%%%%%%%%%%%%%%%%%%%%%%%%%%%%%%%%%%%%%%%%%%%%%%%%%%%%%%%%%%%%

\subsection*{The Monte-Carlo Physarum Machine model}

\begin{figure*}[t!]
\includegraphics[width=0.9\textwidth]{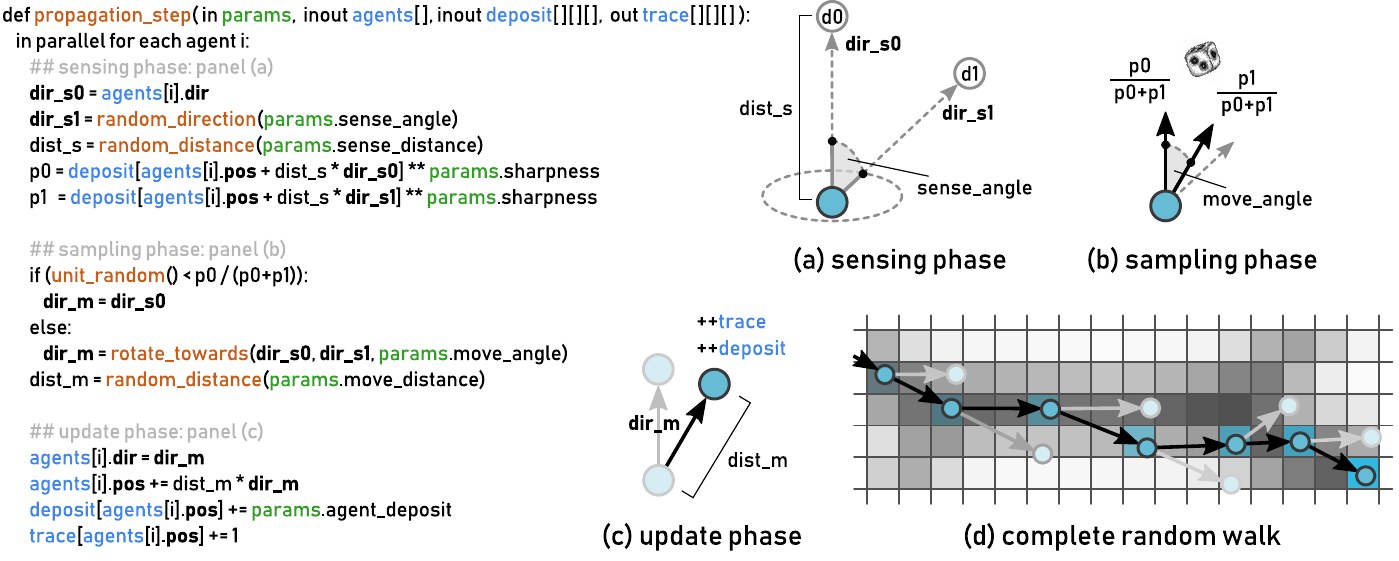}
\caption{\footnotesize
\textbf{Pseudocode and a schematic depiction of the MCPM's agent propagation step.}
In the pseudocode, function calls are highlighted orange, arrays and matrices in blue, structures in green, vectors in boldface.
Each propagation phase is illustrated in the corresponding diagrams (a-c).
Diagram (d) shows a sample random walk of an agent, coupled with the simulation lattice containing the deposit (greyscale) and the trace values (light blue).
We see that the agent's trajectory is a chain of binary directional decisions.
Each propagation step is followed by a relaxation step as explained in the text.}
\label{fig:MCPM_sketch}
\end{figure*}

The employed MCPM model runs in a time-discrete loop, continually producing tentative fits constrained by the input data.
We stop the fitting procedure when the model reaches a soft equilibrium, that is when the solution changes only at the level of process variance.

\begin{figure*}
\includegraphics[width=\textwidth]{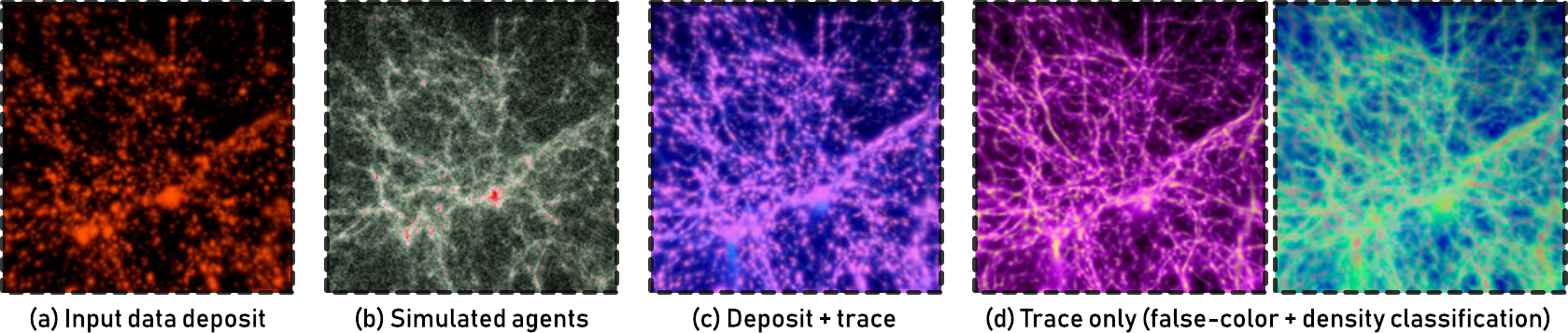}
\caption{\footnotesize
\textbf{Visualizing the elements of the MCPM model.}
(a) Deposit `halos' emitted by the input data.
(b) Particle visualization of the agents (white points) navigating through the input data (red points).
(c) Resulting agents' trace (blue) superimposed over the input deposit halos (purple).
(d) Visualizing only the trace using the same color definitions as in \protect{Figure~\ref{fig:teaser_bp}}.
}
\label{fig:pipeline}
\end{figure*}

The core of MCPM consists of two alternating steps: \textit{propagation} and \textit{relaxation}.
Being a hybrid model, each step is centered around one of the two MCPM's modalities, as follows.
\begin{enumerate}
    \item The \textbf{propagation step} (detailed in Figure~\ref{fig:MCPM_sketch}) iterates over a discrete ensemble of agents, which are the model's device for exploring the simulation domain.
    Each agent's state is represented by a position and a movement direction, which are updated in parallel to navigate through the \textit{deposit field} (see Figure~\ref{fig:pipeline}a).
    The deposit field is stored as a piecewise-continuous 3D lattice of scalar values, representing both the input data as well as the agents' spatial density.
    The deposit effectively guides the agents, as these are propagated with higher likelihood to the regions where deposit values are greater.
    In addition to the deposit, we also maintain a scalar \textit{trace field} (Figure~\ref{fig:pipeline}d) which serves only to record the the agents' spatio-temporal density, but does not participate in their guiding.
    \item The \textbf{relaxation step} ensures that the simulation eventually reaches an equilibrium.
    The deposit field is diffused spatially by a small uniform kernel and decays -- that is, each cell is multiplied by a value $<1$.
    The trace field also decays but does not diffuse in order to preserve features in the estimate of the agents' distribution, which the trace represents.
\end{enumerate}

The input to MCPM is a set of weighted points representing either galaxies, for the SDSS dataset, or dark matter halos, for the Bolshoi-Planck simulation.
These points are static and---just like the agents---emit and inject an amount of deposit at each simulation step, but at the same points in space.
Unlike the agents however, the data deposit is not constant but proportional to each input object's mass.
Since we typically use 5--50 times more agents than input data points (see `Application of MCPM' below), the deposit emitted by the agents is also significantly smaller (subject to fine-tuning).

MCPM's main output is the trace field, which we use in all our subsequent analysis as the indicator of Cosmic Web density.
The trace effectively superimposes the trajectories of all agents throughout the recent fitting history, therefore representing their time-amortized density in space.
Please refer to Figure~\ref{fig:MCPM_sketch} for a schematic depiction of both deposit and trace, and to Figure~\ref{fig:pipeline} for 3D visualizations of the involved quantities.

%%%%%%%%%%%%%%%%%%%%%%%%%%%%%%%%%%%%%%%%%%%%%%%%%%%%%%%%%%%%%%%%%%

\subsection*{Relation of MCPM to previous work}

\begin{figure*}
\includegraphics[width=\textwidth]{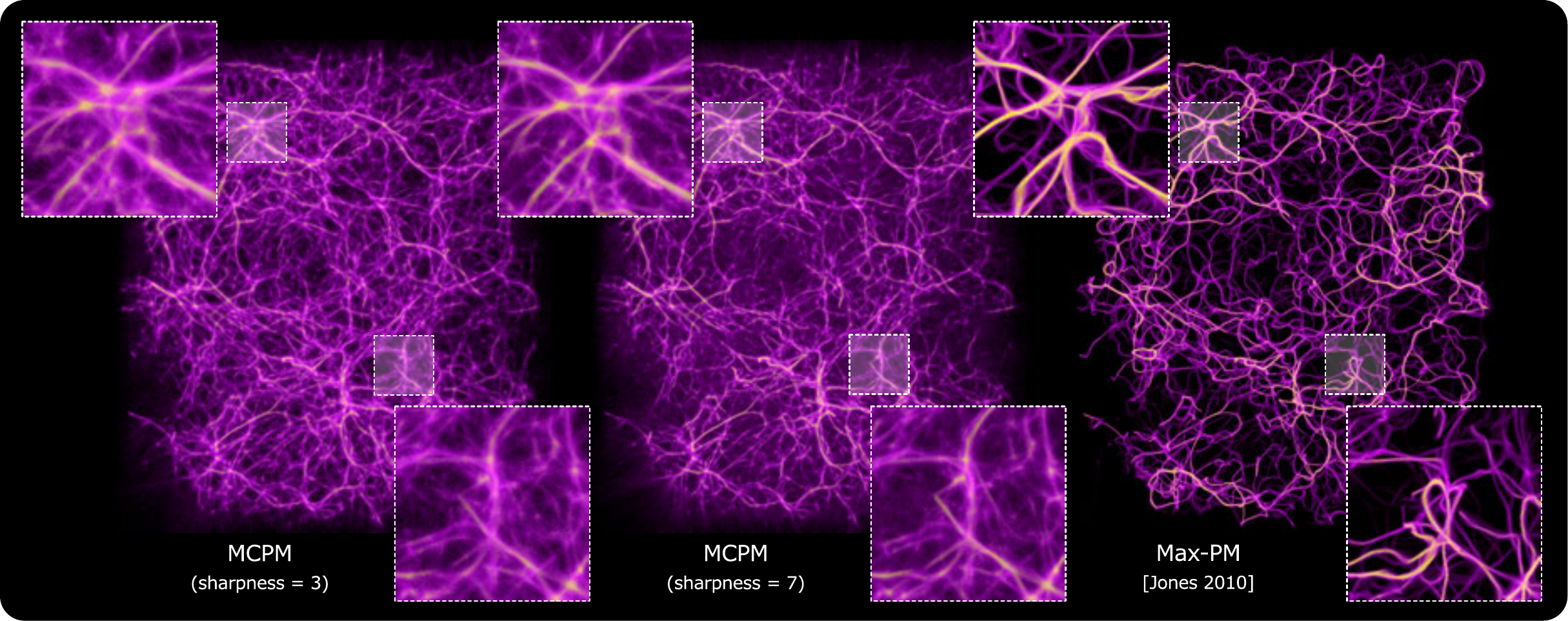}
\caption{\footnotesize
\textbf{Comparison of our MCPM model to Jones's Max-PM model in a 30-Mpc slice from the Bolshoi-Planck dataset.}
Our Max-PM adaptation to 3D uses 9 directional probes for the deposit sampling \citep[cf.][]{Jones:2010aa}, with all other parameters kept identical with MCPM.
Different \texttt{sharpness} values in MCPM trade filament definition for structural complexity (left vs. center);
Max-PM (right) being the limit case with sharp (nearly `tubular') filaments, but lowest geometric complexity and connectivity at filament intersections.
That in turn causes about 5\text{\%} of input halos to be missed by the Max-PM fit, compared to the negligible 0.05\text{\%} of MCPM.
}
\label{fig:model_comparison}
\end{figure*}

\begin{table*}
\small
\centering
\begin{tabular}{r|lcr|l}
    \verb|sense_angle| & $20~[\mathrm{deg}]$ & \quad &
    \verb|sharpness| & $3$--$7$ \\
    \verb|sense_distance| & $2.5~[\mathrm{Mpc}]$ & \quad &
    \verb|agent_deposit| & $0.005$--$0.05$ \\
    \verb|move_angle| & $10~[\mathrm{deg}]$ & \quad &
    \verb|data_deposit| & $\msuneq\ / 10^{11}$ \\
    \verb|move_distance| & $0.2~[\mathrm{Mpc}]$ \\
\end{tabular}
\caption{
\textbf{Parameter values used in the MPCM fitting.}
}
\label{tab:MPCM_params}
\end{table*}

\begin{figure*}
\includegraphics[width=\textwidth]{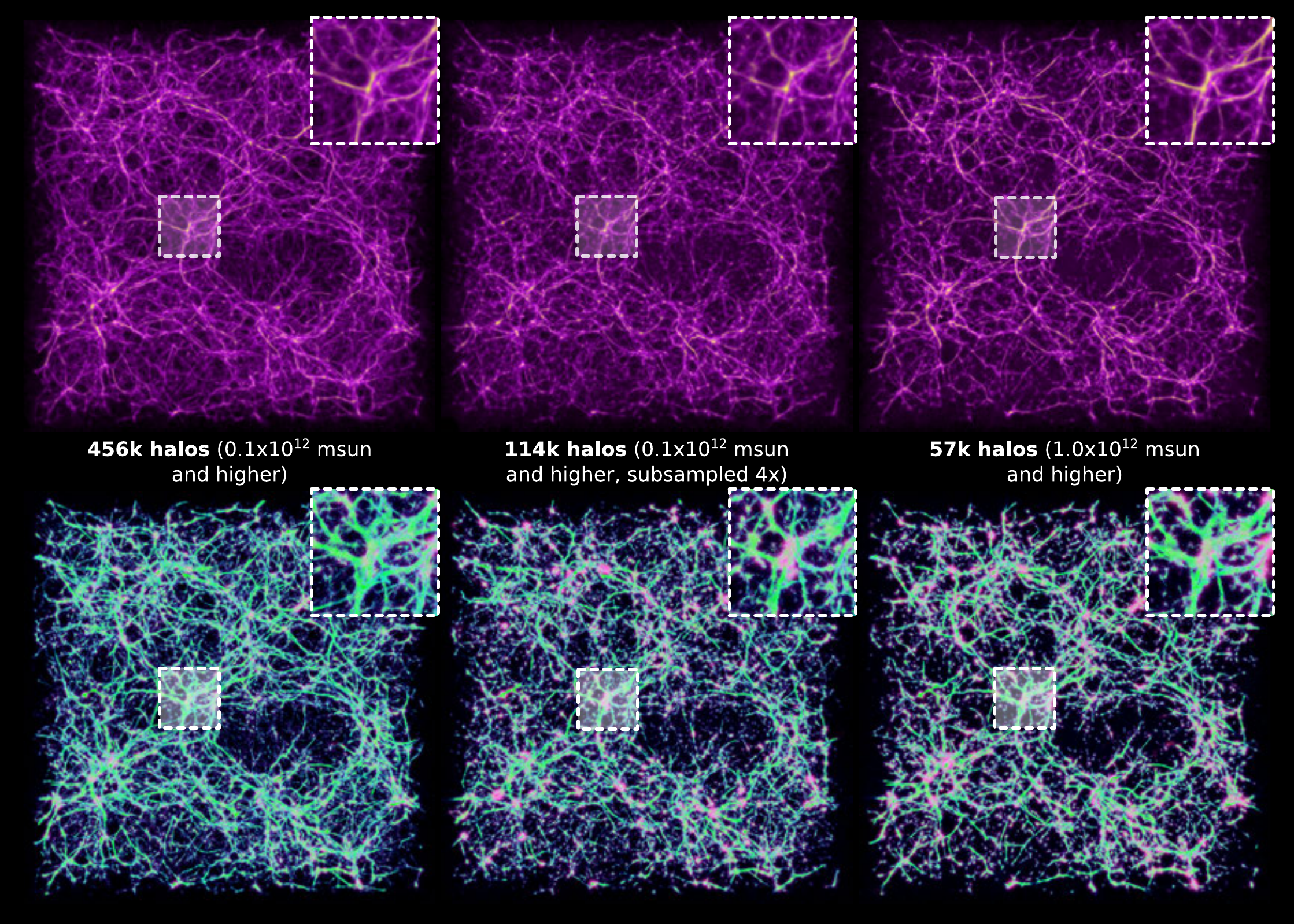}
\caption{\footnotesize
\textbf{Robustness evaluation for the MCPM model.}
We compare fits starting from the 456k most massive halos ($\geq10^{11}$ \msun{}) from the Bolshoi-Planck dataset in the left column.
We apply two decimation schemes: uniform subsampling to a halo \textit{density} comparable to our SDSS dataset (middle column), and selective subsampling to a comparable halo \textit{count} using only halos more massive than $10^{12}$ \msun{} (right column).
In both reduced cases some fine-level details are lost, however the overall structure is well preserved (see sample insets).
}
\label{fig:model_scaling}
\end{figure*}

Being based on the virtual Physarum machine model of \citet{Jones:2010aa}, MCPM shares a number of its features and we encourage the reader to refer to this seminal paper and complement our exposition.
Here we examine the extensions to this model we have made to facilitate our Cosmic Web reconstruction.
\begin{itemize}
    \item \textbf{Generalization from 2D to 3D.}
    The extension of Jones's model to 3D follows with minor modifications, although the obtained network morphologies are quite different due to the peculiar topological properties of 3D space (as shown in Figure~\ref{fig:model_comparison}).
    The only notable modification is the directional sampling (see Figure~\ref{fig:MCPM_sketch}a and the \verb|random_direction| function in the pseudocode therein): we now need to sample a cone of directions rather than simply examine two directions at fixed angles from the agent's current propagation direction.
    
    \item \textbf{Stochastic Monte-Carlo exploration of the domain.}
    Jones's agents \citep{Jones:2010aa} always navigate towards the local maxima of the deposit (thus we will be referring to this model as Max-PM).
    This yields very detailed solutions (example shown in Figure~\ref{fig:model_comparison},\,right), but results in a highly nonlinear behavior especially when the fitting constraints are complex distributions of many points (we typically fit to $10^4$--$10^5$ input points using millions of agents).

    Instead, we view the simulation as a \textit{stochastic process}: in the propagation step the deposit is interpreted as a probability distribution that guides the agents.
    This has multiple implications illustrated in Figure~\ref{fig:MCPM_sketch}.
    First, rather than using fixed distances and directions in the propagation step, we generate them probabilistically: drawing uniformly from a cone for directional sampling (function \verb|random_direction|) and drawing from the 1D Maxwell-Boltzmann distribution for distance sampling (function \verb|random_distance|).
    Second, rather than probing the space of directions more densely \citep{Jones:2010aa} we rely on a single \textit{binary decision}.
    That is (using the \verb|unit_random| function, Figure~\ref{fig:MCPM_sketch}b) we simply choose between mutating or not-mutating the propagation direction.
    This design is commonly used e.g.\ in Markov-chain Monte Carlo methods.
    
    In order to convert the deposit values to sampling probabilities (\verb|p0| and \verb|p1| in the pseudocode) we opted for a geometric proportionality controlled by a single \verb|sharpness| parameter $\in \mathbb{R}^{+}$.
    As demonstrated in Figure~\ref{fig:model_comparison}, this grants direct control over the actual sharpness (acuity, or falloff) of the filamentary structures: for $\verb|sharpness| \rightarrow 0$ the agents diffuse uniformly and all anisotropic structures gradually disappear; to the other extreme, the model slowly converges to the Max-PM behaviour \citep{Jones:2010aa} as $\verb|sharpness| \rightarrow \infty$.
    
    \item \textbf{Extraction of the agent trace.}
    While the deposit field could serve as an indicator function for the cosmic web density reconstruction, this would not yield optimal results because the relaxation step effectively acts as a low-pass filter.
    Instead, we introduce the trace field, which is a direct recording of the temporally averaged (aggregate) agent trajectories.
    Thanks to the stochastic formulation of our model, the trace is equivalent to a probability density function (after proper normalization), and serves as a suitable basis for constructing mappings to the cosmic overdensity and other relevant cosmological quantities.
\end{itemize}

%%%%%%%%%%%%%%%%%%%%%%%%%%%%%%%%%%%%%%%%%%%%%%%%%%%%%%%%%%%%%%%%%%

\subsection*{Application of the MCPM model}

Similarly to the actual \textit{Physarum polycephalum} organism, MCPM and its other virtual realizations produce a set of pathways that---with respect to a relevant measure in the operating domain---form approximate optimal transport networks spanning the input data (`food').
The conjecture that gravitationally relaxed matter has the tendency to form such networks is indirectly supported by our findings.

\begin{figure}

    \centering
    \includegraphics[width=\linewidth]{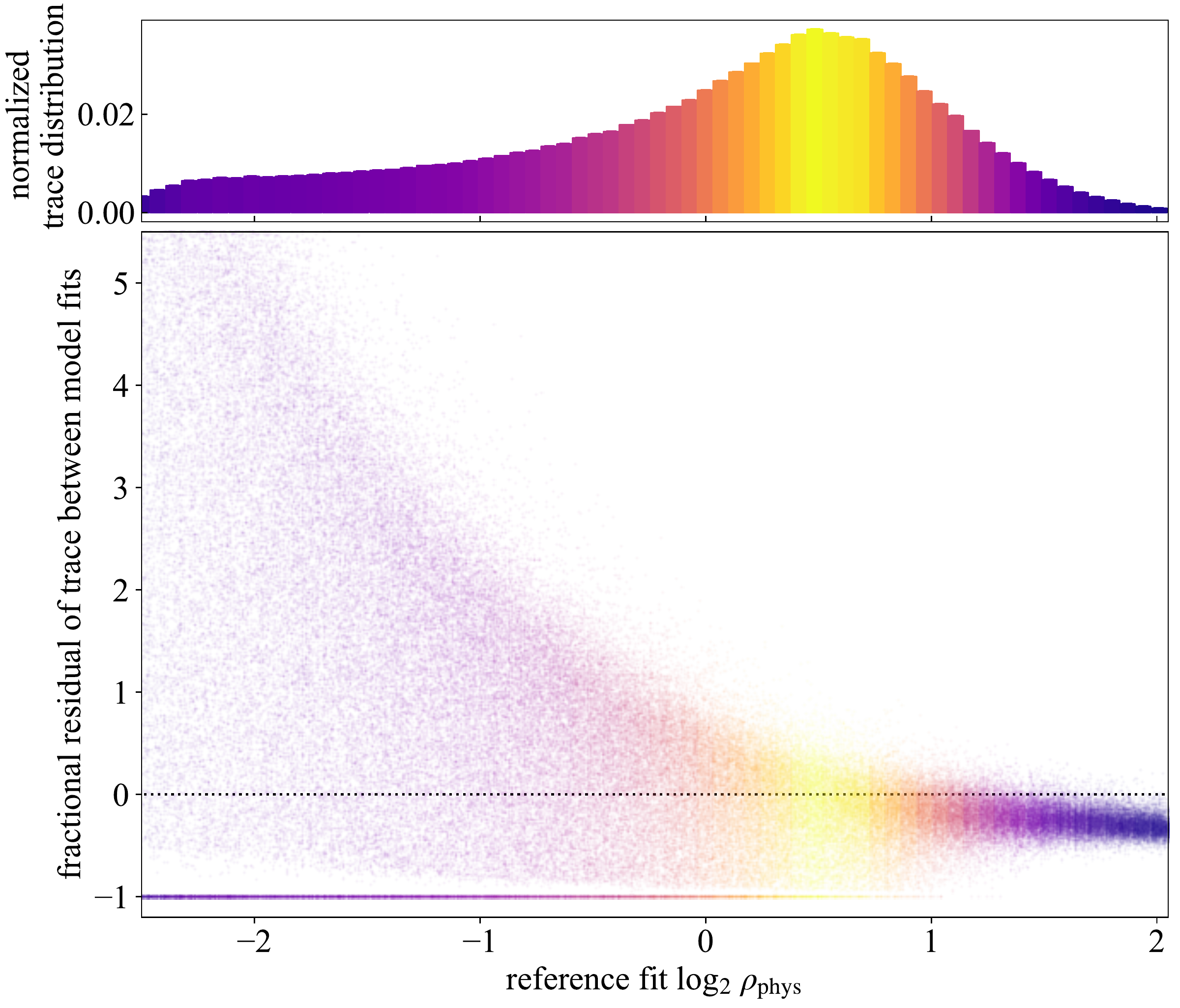}
    \caption{Comparing the model trace values for MCPM fits to Bolshoi-Planck catalogs including halos with  $M > 10^{11} \msuneq$ (reference fit) and with $M > 10^{12}\msuneq$ (comparison fit).  The bottom panel shows residuals between standardized trace values (\rhophys{}) found in the comparison and reference fits ([comparison-reference]/reference) at the same spatial locations.  The upper histogram shows the distribution of \rhophys{} in the reference fit.  Included in the bottom panel are points sampled evenly in bins of \rhophys{} to highlight the relationship across the full dynamic range, and both panels are color coded to emphasize the relative frequency of points in each \rhophys{} bin.  We note agreement between the models near the peak of the distribution, while the downsampled model fit diverges in the lower tail of the distribution.  }
     \label{fig:downsampleFitResid}
\end{figure}

\begin{figure*}[t!]
\centering
\includegraphics[width=0.9\textwidth]{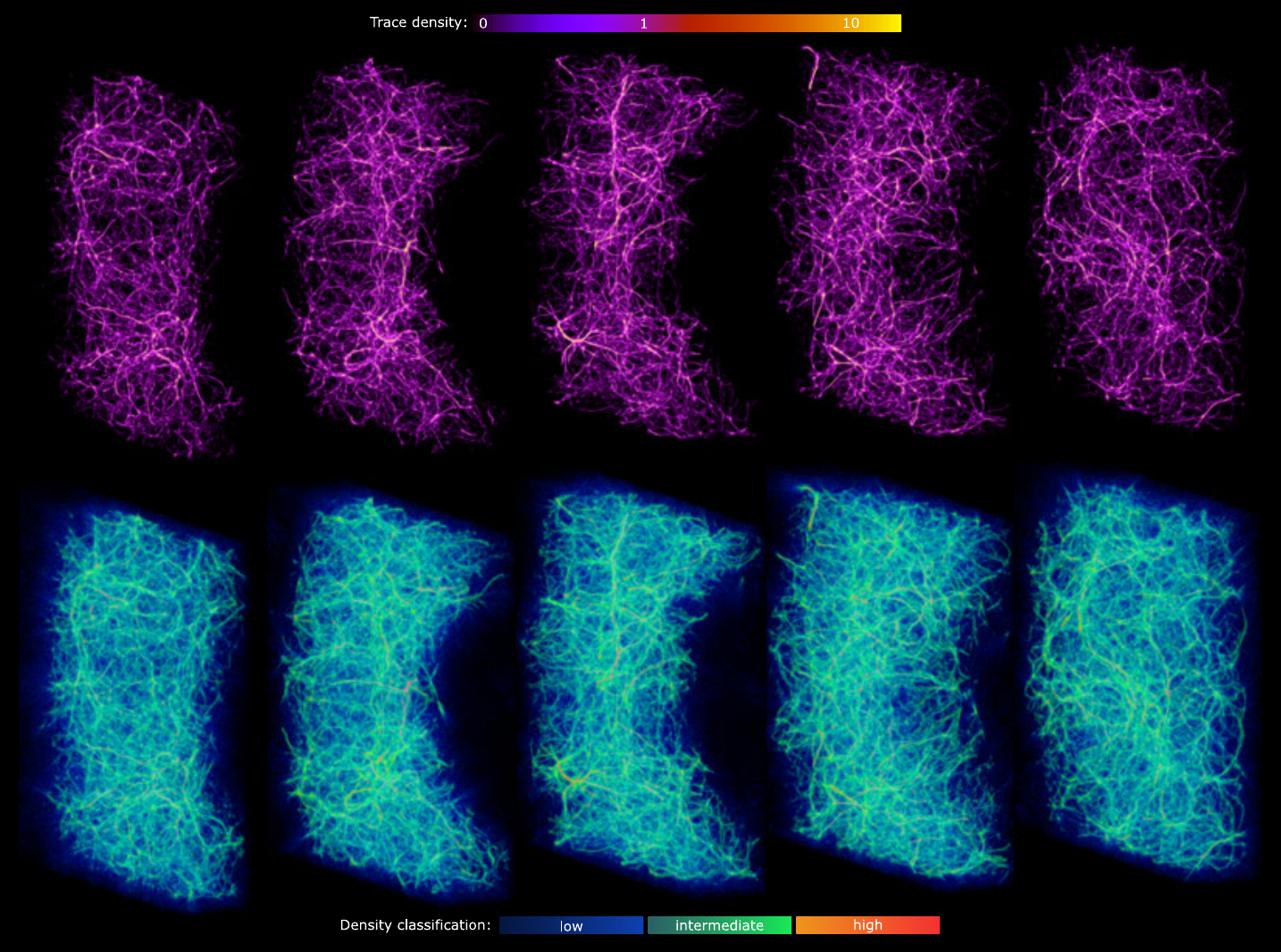}
\caption{\footnotesize
\textbf{Slicing through the density field reconstructed from the most massive 456k halos ($\geq10^{11}$ \msun) of the Bolshoi-Planck simulation volume.}
Top: False-color visualization of the mapped overdensity.
Bottom: Spatial distribution of the trace density values, divided into three intervals.}
\label{fig:teaser_bp}
\end{figure*}

\begin{figure}
\centering
    \vspace{-2mm}
    \includegraphics[width=0.5\textwidth]{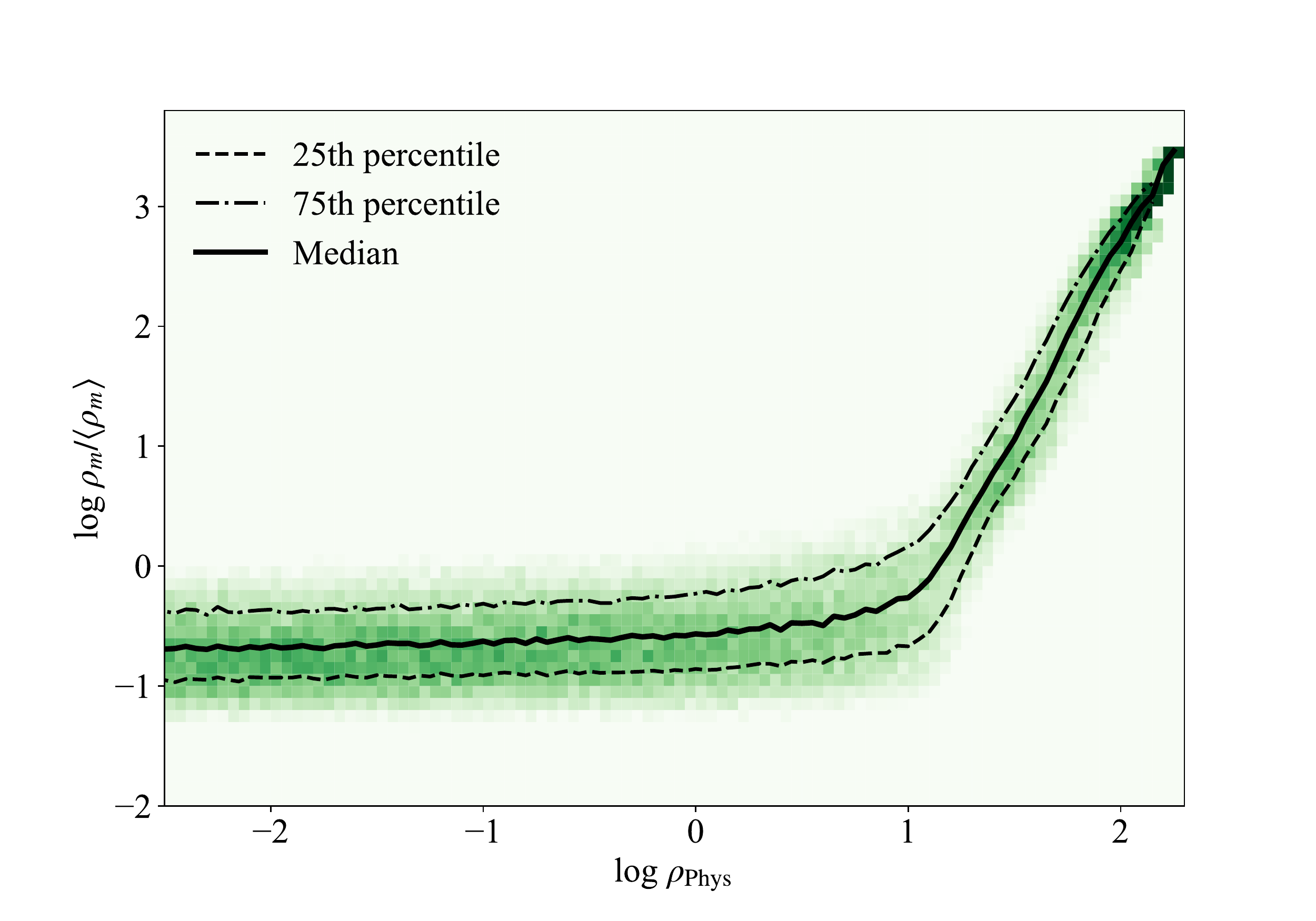}
    \hfill
    \vspace{-5mm}
    \caption{\footnotesize \textbf{Calibration of the Physarum densities with dark matter densities from the Bolshoi-Planck simulation.}  Within bins of MCPM trace values ($\rho_{\rm Phys}$) from our fit to the Bolshoi-Planck halo catalog, we matched the spatial locations of points in the MCPM volume to local density values from the Bolshoi-Planck particle data.  This provides a mapping to the cosmic mean matter density (\rhorhom).  The bold black curve shows the median $\rho/\rho_m$ values in each bin, and the dashed and dash-dotted black curves denote the 25th and 75th percentiles, respectively, of each \rhophys\ bin's distribution.  \label{fig:calibration}}  
\end{figure}

Our implementation of MCPM is written in C++ and uses GPU compute shaders for executing the parallel propagation and relaxation steps, as well as the volume visualization.
The GPU implementation allows us to run the fitting with 2M agents over $960^3$ deposit and trace grids at interactive speeds, using an NVIDIA TitanX GPU.
The model typically converges in fewer than 1000 timesteps (about 1 minute).
All of the model's meta-parameters (see pseudocode in Figure~\ref{fig:MCPM_sketch}, and Table~\ref{tab:MPCM_params} for typical values) can be modified at any point during the simulation, which allows for easy tuning of the obtained 3D structures.

Of critical importance is the question regarding MCPM's \textit{robustness} when applied to datasets with different spatial densities.
This will become particularly important when applying MCPM to galaxy samples at higher redshift in flux-limited surveys, such as the SDSS, which will contain only more massive, rarer galaxies.
We therefore examine this robustness in Figure~\ref{fig:model_scaling}, where we fit the MCPM model to three subsamples of the Bolshoi-Planck halo catalog: including all $M > 10^{11} \msuneq$ halos ($\sim$456,000 objects), including all $M > 10^{12} \msuneq$ halos ($\sim$57,000 objects), and randomly subsampling the $M > 10^{11} \msuneq$ catalog by a factor of 1/4 ($\sim$114,000 objects).  For this comparison, we simply adopted the same MCPM parameters for all fits without additional optimization.  Figure~\ref{fig:model_scaling} demonstrates that MCPM finds salient filamentary structures even when the input data is significantly reduced.
Crucially, if configured well, the model does not \textit{imagine} nonexistent structures: if the input data is too sparse with regard to the characteristic feature scales (per given geometric configuration), the agents will either be near-uniformly dispersed in space, or form only `fuzzy' tentative (low-probability) filaments.
This is another desirable property of MCPM relative to the Max-PM model \citep{Jones:2010aa}, which has no way to assess the `confidence' in the reconstructed structures.

We more quantitatively assess the impact of sparsely sampled data by comparing the trace values from the mass-downsampled fit shown in Figure~\ref{fig:model_scaling} (right) with the more densely sampled fit (left; hereafter referred to as the reference fit).  Upon extracting trace volumes from each fit, we standardized each distribution by subtracting its mean and dividing by its standard deviation.  We then sampled the reference fit in 110 bins of $\Delta$log$_2$ \rhophys $\sim$ 0.07, identifying 2000 cells with trace values in each bin, and then logged the trace values at identical locations in the comparison fit.  Figure \ref{fig:downsampleFitResid} shows the fractional residuals ([comparison-reference]/reference) as a function of reference fit trace (bottom panel) as well as the global reference trace distribution (top panel).  Even without optimizing the fits for the downsampled dataset, we note that the reference and comparison fits agree well at the peak of the \rhophys\ distribution.  Within the lower tail of the distribution, the two fits diverge from one another, suggesting that including only more massive halos will poorly constrain low-density structures.  This result is not surprising, although we underscore that the comparison fit was not tuned even given the dramatic decrease (by a factor of 1/8) in datapoints used in the fit.  A more rigorous performance and optimization analysis of the model under varying conditions (including sampling density) will be presented in a subsequent publication.

%%%%%%%%%%%%%%%%%%%%%%%%%%%%%%%%%%%%%%%%%%%%%%%%%%%%%%%%%%%%%%%%%%

\subsection*{MCPM trace/Bolshoi-Planck matter density calibration}
The `trace' output of the Physarum model effectively provides density values (\rhophys) throughout the volume sampled by our SDSS galaxies.  To calibrate these values to more cosmologically meaningful terms of the cosmic matter density (\rhorhom), we fitted the MCPM model to the halo catalog for the Bolshoi-Planck $z=0$ snapshot produced by the Rockstar algorithm \citep{Behroozi:2013aa} (Figure \ref{fig:teaser_bp}).  We then sampled the trace values within this volume in logarithmic bins of $\Delta$log$_2$~\rhophys~$= 0.05$, crossmatching with the cube of smoothed matter density values, and produced a mapping between the two via the running median, as shown in Figure~\ref{fig:calibration}.  

We highlight two distinct features of this mapping: 1) The relatively flat relation for log$_2$ \rhophys $\lesssim 1$ is a result of the larger dynamic range spanned by MCPM trace values than matter density values of the Bolshoi-Planck data product.  2) The MCPM trace and BP matter densities are well correlated for log$_2$ \rhophys $\gtrsim 1$.  It is in this regime where the mapping shown on the upper horizontal axis of Figure \ref{fig:absorptionSignal} panels is well matched.  Consequently, the densities just below and including where our detected absorption signal occurs also has the most robust mapping between \rhophys\ and \rhom. Finally, we standardized the overall distributions of \rhophys\ between the MCPM fits of the BP halos and SDSS galaxies.  For each distribution, we calculated its mean and standard deviation then subtracted the mean and divided by the standard deviation.

\subsection*{H I absorption signal}
As briefly sketched in the main text, our MCPM Cosmic Web density reconstruction provides a \rhophys\ value at every point in the 3D volume, which in turn provides a \rhophys\ value for each pixel in the HST/COS spectra probing the volume.  The galaxy survey data we input to MCPM includes the 2D celestial coordinates plus the luminosity distance from the galaxies' redshifts.  Therefore, we mapped coordinates in the output MCPM density field to spectral pixels using the celestial coordinates of each HST/COS sightline and adopted as each pixel's `distance' the luminosity distance at $z_{\rm pix} = \lambda_i/1215.67$\,\AA\ - 1, where $\lambda_i$ is the observed wavelength of each pixel and 1215.67\,\AA\ is the rest wavelength of \hone\ \lya.  We then tabulated the normalized flux \rhophys\ values for each pixel in each of the 346 sightlines in our sample.

The observed wavelength range for \hone\ \lya\ corresponding to the redshifts of our galaxy sample is $1231.25-1257.96$\,\AA, which includes several transitions observed in the Milky Way's interstellar medium (ISM): \textsc{N~v} $\lambda\lambda$ 1238, 1242\,\AA; \textsc{Mg~ii}  $\lambda\lambda$ 1239, 1240\,\AA; and \textsc{S~ii} $\lambda\lambda$ 1250, 1253\,\AA. To avoid introducing a spurious signal due to absorption from these transitions, we masked out the following spectral regions from our analysis (all wavelengths in \AA): (1238.00, 1240.60), (1242.10, 1243.20), (1250.20, 1250.95), and (1253.30, 1254.30).  

We then calculated the median normalized flux of the spectral pixels in bins of $\Delta$log$_2$~\rhophys~$=~0.4$ (prior to the standardization step described above).  We adopted the standard deviation of medians from 1000-pixel bootstrap resamples as the uncertainty on the median.  Although we had masked spectral regions likely impacted by the absorption from the Milky Way ISM, a residual, spurious absorption signal certainly remains from interloping absorbers at  redshifts $z>0$.  This signal was apparent upon our initial calculation of the binned median fluxes, particularly at low densities (log$_2$ \rhophys $< 0.8$), where the median normalized flux remained relatively flat with increasing density but was offset from unity.  We verified that the offset was indeed due to interloping absorption by isolating a subset of the sightlines for which we had identified the  absorption lines at all redshifts.  We then calculated the median flux for the sightlines in this subset with known interloping absorption lines having observed wavelength within the range analyzed and for those sightlines without such interlopers.  Indeed, upon comparing the absorption profiles, the residual signal arose significantly in sightlines with interloping absorbers but did not in those free of interlopers.  In the full dataset, we corrected for this spurious absorption signal as follows. To establish the baseline flux $f_b$, we performed a linear least-squares fit to the median flux values at log$_2$ \rhophys $< 0.8$.  The fitted slope was 0.00026 $\pm$ 0.00097, suggesting no statistically significant change in the signal over this range.  We adopted the y-intercept ($f = 0.981$) of the fit as $f_b$ in Equation 1. The offset of this baseline from a normalized \lya\ flux of 1 (i.e., no \hone\ absorption) is mostly due to interloping lines from other redshifts.

\begin{figure}
\centering
    \includegraphics[width=\linewidth]{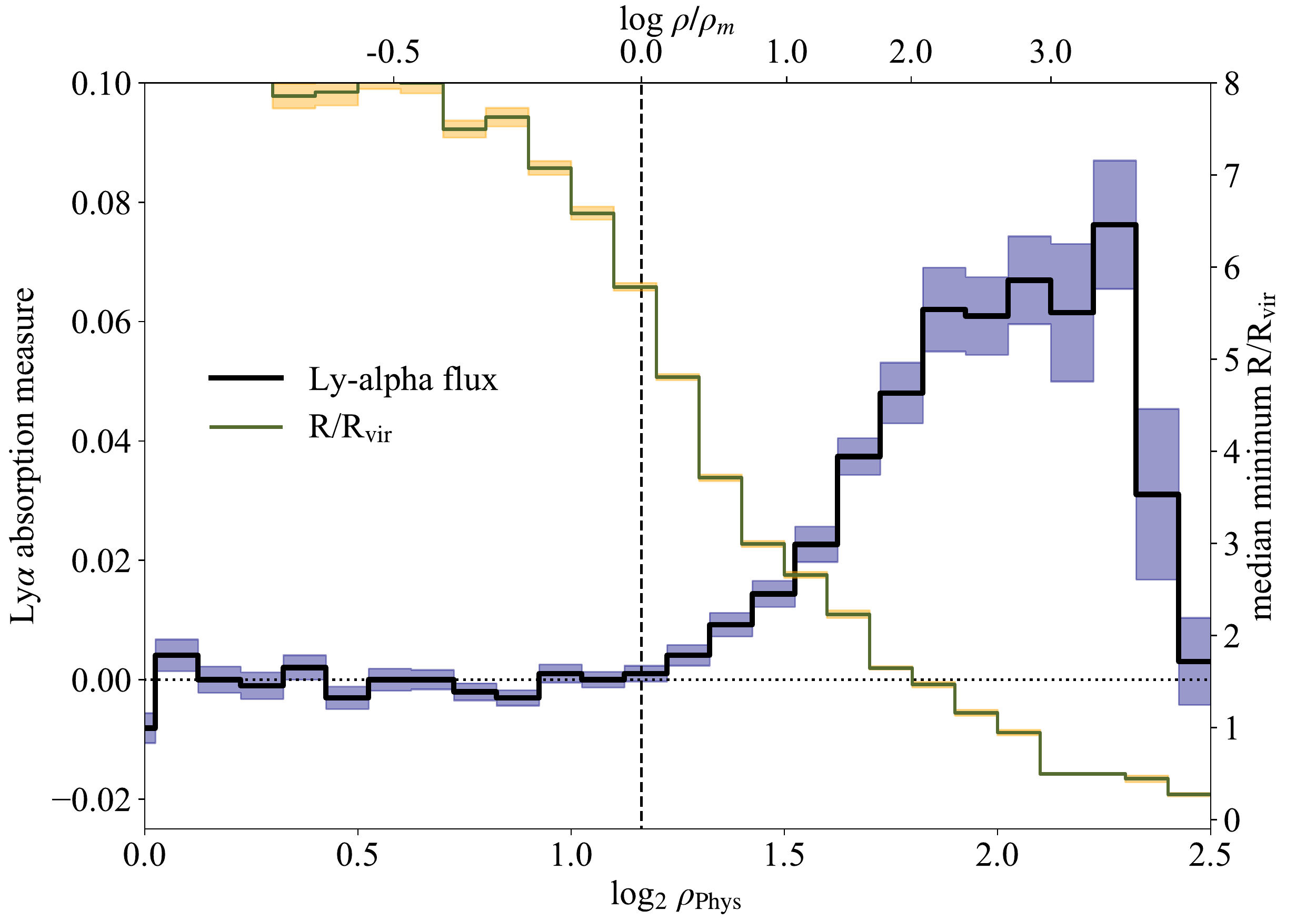}

    \caption{\footnotesize \textbf{Density-dependent absorption signal juxtaposed with nearby galaxy statistics} As in
    Figure~\ref{fig:absorptionSignal}, the black curve and blue shading shows the absorption measure as a function of environmental density.  The green curve shows the median of the impact parameters normalized to the virial radii of closest galaxies to the sightline pixels represented in each density bins (values are shown in the right-hand vertical axis).  A vertical dashed line marks the cosmic mean matter density and the approximate beginning of the intermediate-density regime.} \label{fig:absorptionSignalGalaxyRho}

\end{figure}

To better understand the galaxy environments probed by our absorption data, we crossmatched each spectral pixel with the SDSS galaxy catalog, searching for galaxies within 5\,Mpc of the sightline and having redshifts within $\pm 500$\,\kms of $z_{\rm pix}$.  For each pixel, we found the `nearest' galaxy according to two criteria: 1) that with the smallest impact parameter in proper distance and 2) that with the smallest impact parameter normalized to the galaxy's virial radius \citep{Burchett:2016aa}.  In the lowest \rhophys\ bins (log$_2$ \rhophys $<1.5$), no galaxies were found within 5\,Mpc for a large fraction of the pixels, as expected.  Similar to the normalized flux, we then calculated the medians and uncertainties of the smallest impact parameters in each \rhophys\ bin.  Figure~\ref{fig:absorptionSignalGalaxyRho} shows, in addition to the absorption signal from Figure \ref{fig:absorptionSignal}, the median of the impact parameter to the nearest galaxy within each bin (right vertical axis), where the impact parameters have been normalized by galaxy virial radius (R$_{\rm vir}$).  As we alluded in the main text, the high-density regime begins where the spectra typically probe within $\sim$ 1-2\,R$_{\rm vir}$, whereas the intermediate-density regime begins beyond $\sim$ 4-5\,R$_{\rm vir}$ of any galaxy.

%\clearpage

\bibliographystyle{apj}

\end{document}